\documentclass[12pt,english,aps,prb,reprint,amsmath,amssymb,onecolumn,notitlepage]{revtex4-2}
\usepackage[T1]{fontenc}
\usepackage[latin9]{inputenc}
\usepackage{color}
\usepackage{babel}
\usepackage{bm}
\usepackage{amsmath}
\usepackage{amssymb}
\usepackage{graphicx}
\usepackage{microtype}
\usepackage[unicode=true,pdfusetitle,
 bookmarks=true,bookmarksnumbered=true,bookmarksopen=true,bookmarksopenlevel=1,
 breaklinks=false,pdfborder={0 0 0},pdfborderstyle={},backref=false,colorlinks=true]
 {hyperref}

\makeatletter

\providecommand{\tabularnewline}{\\}

\usepackage{braket}
\usepackage{mathtools}
\usepackage{graphicx}

\usepackage{microtype}
\hypersetup{linkcolor = blue,
            urlcolor  = blue,
            citecolor = blue,
            anchorcolor = blue}

\makeatother

\begin{document}
\title{Unique determination of localized basis in molecular spin}
\author{Le Tuan Anh Ho}
\email{chmhlta@nus.edu.sg}

\affiliation{Department of Chemistry, National University of Singapore, 3 Science Drive 3 Singapore 117543}
\author{Liviu Ungur}
\email{chmlu@nus.edu.sg }

\affiliation{Department of Chemistry, National University of Singapore, 3 Science Drive 3 Singapore 117543}
\begin{abstract}

Localized basis plays an important role in comprehending the magnetic dynamics in molecular spins from a physics perspective. Nonetheless, the uniqueness and rigor of its determination have received limited attention. In this study, we propose a new determination of the localized basis applicable to both non-Kramers and Kramers molecular spin systems, leveraging the time-reversal symmetry of the spin Hamiltonian and the molecular spin's main magnetic axis. By introducing this, we establish a distinct and practical means of determining the localized basis, enabling the association of a molecular spin wave function with either an \textquotedbl up\textquotedbl{} or \textquotedbl down\textquotedbl{} magnetic moment orientation in molecular spins. This finding facilitates a comprehensive interpretation of magnetic dynamics and simplifies the construction of theoretical models for materials analysis.
\end{abstract}
\maketitle
\global\long\def\hmt{\mathcal{H}}%
\global\long\def\vt#1{\overrightarrow{#1}}%

\global\long\def\chip{\chi'}%
\global\long\def\chipp{\chi''}%

\global\long\def\tn{\mathrm{tn}}%
\global\long\def\vtrm#1{\bm{\mathrm{#1}}}%

\section{Introduction}

Recent advancements in molecular spins have sparked considerable interest due to their potential applications in spintronics \citep{Bogani2008,Mannini2010,Clemente-Juan2012,Coronado2019}, high-density storage \citep{Guo2018,Goodwin2017,Woodruff2013d,Bartolome2017}, and quantum computing \citep{Leuenberger2001,Gaita-Arino2019,Leuenberger2003,Moreno-Pineda2018,Carretta2021}. While current research focuses on dynamic properties such as decoherence and magnetization relaxation \citep{Escalera-Moreno2020,Lunghi2020a,Garlatti2021a,Serrano2020,Wang2020,Luo2021}, there is a growing need to understand the underlying mechanisms and gain deeper physical insights. To achieve this, an effective approach is to describe magnetic dynamics using the master equation of spin density matrix in a localized basis \citep{Garanin2011,Leuenberger2000,Ho2017,Blum1996,Gatteschi2007}, where basis wave functions are approximately localized in either the left or right potential well ($x$-axis is the magnetic quantum number $m$) \citep{Gatteschi2003,Chudnovsky2005,Villain2000,Gatteschi2007,Leuenberger2000}. This allows for an easy visualization of the variation of the spin magnetic moment orientation, facilitating a comprehensive understanding of magnetic dynamics and its associated mechanisms. 

Existing literature on localized basis determination, as far as we are aware, reveals two main approaches. The first approach \citep{Gatteschi2007} constructs localized basis states corresponding to a doublet of the molecular spin from an equally linear combination of two normalized eigenstates of the doublet at resonance. However, this approach neglects phase factors associated with eigenstates and also fails to differentiate between non-Kramers and Kramers doublets. The Kramers degeneracy in the latter introduces additional degrees of freedom in the choice of eigenstates, compromising the uniqueness of localized basis determination of this approach. Consequently, different choices of localized basis may lead to varying interpretations of magnetic dynamics and physical mechanisms.

The second approach, utilized in the quantum chemistry software MOLCAS \citep{Aquilante2016,Chibotaru2012b,Ungur2011a}, determines localized basis states by treating each doublet as pseudospins 1/2 and diagonalizing the submatrix of the magnetic moment component along the doublet's main magnetic axis. This approach ensures the uniqueness of the basis states corresponding to a specific doublet. However, since the energy spectrum of a molecular spin in general consists of multiple (quasi-) doublets and their main magnetic axes are typically not of the same orientation, the process of determination of the localized basis states corresponding to different doublets cannot be done simultaneously with only a specific $z$-axis. As a result, localized basis states determined for one doublet may yield non-localized basis states for other doublets.

The limitations of previous approaches \citep{Gatteschi2007,Aquilante2016,Chibotaru2012b} necessitate the development of a rigorous and practical determination of the localized basis. This determination should both ensure the uniqueness of the localized basis and provide a detailed and clear procedure. In light of these requirements, our work proposes an unequivocal determination of localized basis for molecular spin systems, based on the time-reversal symmetry and the main magnetic axes of the whole molecular spin.

\section{Determination of the localized basis for non-Kramers and Kramer system}

\begin{figure}
\centering{}%
\begin{tabular}[t]{cccc}
a) &  & b) & \tabularnewline
 & \includegraphics[width=0.4\textwidth]{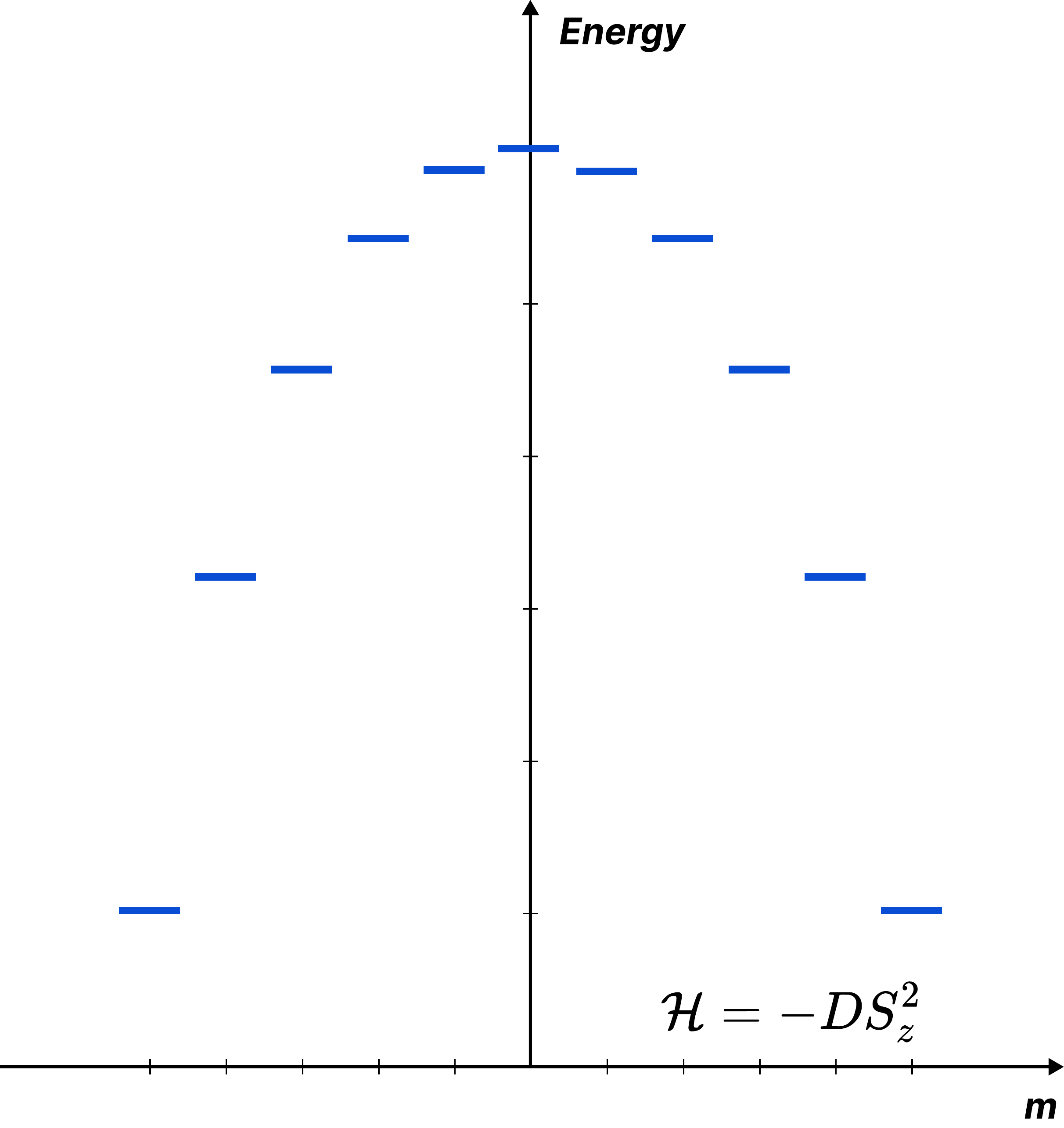} &  & \includegraphics[width=0.4\textwidth]{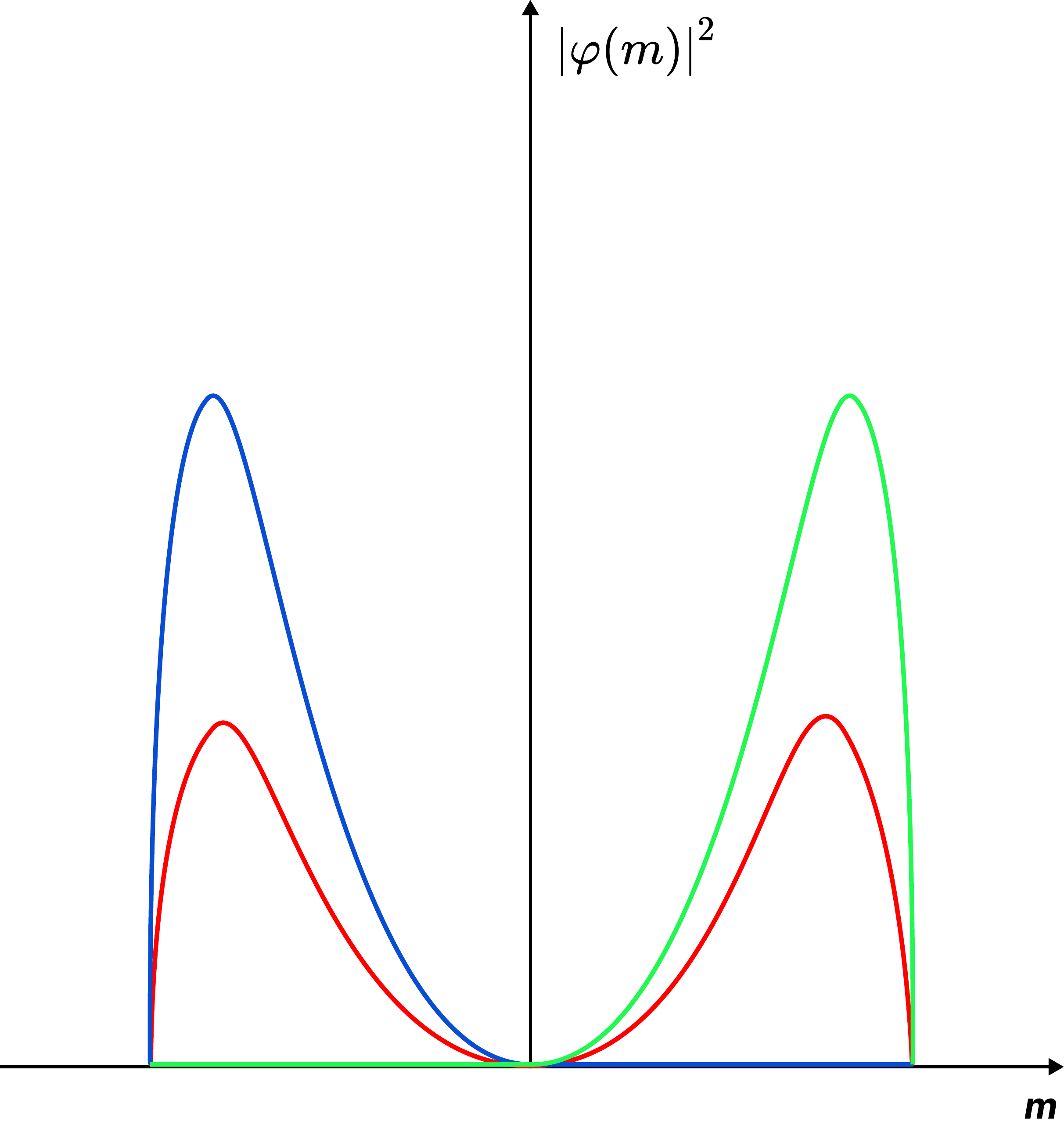}\tabularnewline
\end{tabular}\caption{a) Schematic representation of the energy spectrum of a molecular spin system as a function of the magnetic quantum number $m$. b) Schematic representation of the probabilities $\left|\varphi\left(m\right)\right|^{2}$ as a function of the magnetic quantum number $m$ for localized states (blue and green) and an equally delocalized state (red). \label{fig:Schematic-diagram}}
\end{figure}

A generic molecular spin $S$ consisting of $2S+1$ energy levels will be considered. The spin Hamiltonian describing this molecular spin is formulated in terms of the extended Stevens operators (ESO), $O_{p}^{q}\left(\vtrm S\right)$, as follows: 
\begin{equation}
\hmt=\sum_{p=2,4,6}\sum_{q=-p}^{p}B_{pq}O_{p}^{q}\left(\vtrm S\right),
\end{equation}
where $B_{pq}$ are the associated ESO real-valued coefficients. At zero magnetic field, the energy spectrum of this spin exhibits multiple doublets and/or singlets, as depicted schematically in Fig. \ref{fig:Schematic-diagram}a. 

Let's consider one of these doublets, characterized by its eigenstates $\ket{\psi}$ and clarify the concept of localization within the framework of this study. In the natural basis eigenstates $\ket{m}$ of the longitudinal spin operator $S_{z}$, these can be decomposed as: 
\begin{equation}
\ket{\psi}=\sum\varphi\left(m\right)\ket{m},
\end{equation}
where $\varphi\left(m\right)$ denotes the wave function of the magnetic quantum number $m$. In the scope of our research, localization refers to the confinement or concentration of a wave function $\varphi\left(m\right)$ to a specific region along the magnetic quantum number $m$ $x$-axis. It signifies the tendency of the quantum state to be predominantly found in a particular area or state. To gain a clearer understanding of this concept, we have illustrated various scenarios in Fig. \ref{fig:Schematic-diagram}b where the probabilities $\left|\varphi\left(m\right)\right|^{2}$ are plotted, showcasing cases of localization on the left-hand side, right-hand side, or an equal distribution on both sides.

In principle, a molecular spin doublet can be regarded as a pseudo two-level system. Similar to a two-level system, the choice of the appropriate basis can significantly aid in finding solutions or gaining insights into the dynamic physical processes occurring within the system. Drawing inspiration from the two-level system, where adiabatic or diabatic bases can be interchangeably used depending on the context, we aim to define/determine these bases for the molecular spin doublets.

While it is evident that the adiabatic basis for a molecular spin doublet consists of the doublet's eigenstates, the same does not hold for the diabatic basis. In the two-level system, the diabatic basis is constructed from the eigenstates of the $\sigma_{z}$ of the spin 1/2. These states correspond to the extremum values of the longitudinal magnetic moment of the spin and, to some extent, behave like the ``up'' and ``down'' states of the doublet. For a doublet of the molecular spin with a general quantum number $S$, this leaves only two options for the determination of the diabatic basis states: 1) these states optimize the longitudinal magnetic moment spin $\tilde{\sigma}_{z}$ of the pseudospin 1/2 corresponding to the doublet, or 2) these diabatic basis states optimize the longitudinal magnetic moment spin $S_{z}$ of the whole spin. The former has been utilized in the previously mentioned second approach and we have shown in the introduction that it leads to an impractical determination. Hence, it is logical that within the vector subspace spanned by the doublet, the diabatic basis associated with a molecular spin doublet must optimize the longitudinal magnetic moment spin $S_{z}$ of the whole spin. This optimization of the longitudinal magnetic moment within the doublet subspace will result in the wave function $\varphi\left(m\right)$ of each diabatic basis state becoming localized on the left-hand side or right-hand side (the $x$-axis is the magnetic quantum number $m$). To maintain consistency with the terminology used in the first approach \citep{Gatteschi2007} and other works in single-molecule magnetism, we refer to these diabatic basis states as localized basis states to emphasize the localization of the corresponding wave function.

In molecular magnetism, a localized basis corresponding to a doublet serves a role similar to that of a diabatic basis in a two-level system. In addition to sharing advantages with the diabatic basis in a two-level system, the localized basis offers an intuitive means to analyze the magnetization relaxation processes through changes in the orientation of the magnetic moments of a population (or equivalently, changes in population corresponding to specific orientations of the magnetic moment). This aids in identifying the dominant relaxation pathway for magnetization. Furthermore, by utilizing the localized basis rather than the basis of energy eigenstates, it becomes possible to quantify quantum tunneling of magnetization, a crucial relaxation process in molecular magnetism \citep{Garanin1997,Leuenberger2000,Ho2017}. 

Having recognized the advantages of the localized basis, our scientific objective in this work is to address the question of how to uniquely determine this distinctive localized basis for both non-Kramers and Kramers molecular spin systems. In addition to rationalizing our determination method, we will also analyze the distinguishing features of our approach in comparison to others. For clarity, hereinafter we denote the eigenstates of the spin Hamiltonian corresponding to the $m^{\mathrm{th}}$ doublet as $\ket{\psi_{m}}$ and $\ket{\psi_{m'}}$. We also use $\theta$ to represent the time-reversal operator.

\subsection{Non-Kramers system}

As there is no Kramers degeneracy in this kind of system, the $m^{\mathrm{th}}$ doublet is in fact a quasi-doublet. Under zero magnetic field, eigenstates $\ket{\psi_{m}}$ and $\ket{\psi_{m'}}$ have approximately equal eigenenergies $\varepsilon_{m}$ with an energy gap $\Delta_{m}$: 
\begin{align}
\hmt_{0}\ket{\psi_{m}} & =\left(\varepsilon_{m}+\frac{\Delta_{m}}{2}\right)\ket{\psi_{m}},\\
\hmt_{0}\ket{\psi_{m'}} & =\left(\varepsilon_{m}-\frac{\Delta_{m}}{2}\right)\ket{\psi_{m'}},
\end{align}
where $\Delta_{m}$ is called the tunnel splitting. Invariance of the spin Hamiltonian under time-reversal operator implies that $\theta\ket{\psi_{m}}$ and $\theta\ket{\psi_{m'}}$ are proportional to $\ket{\psi_{m}}$ and $\ket{\psi_{m'}}$, respectively. Furthermore, it has been widely known that the eigenstates of non-Kramers system can always be chosen to be real in the sense that $\theta\ket{\psi_{n}}=\pm\ket{\psi_{n}}$ \citep{Griffith1963a,Chudnovsky2005}. Indeed, given an arbitrary eigenstate $\ket{\psi}$ of $\hmt_{0}$, we can always make it real by transforming it as follows: 
\begin{align}
\ket{\psi_{\pm}} & =\ket{\psi}\pm\theta\ket{\psi}.
\end{align}
It is obvious that $\theta\ket{\psi_{\pm}}=\pm\ket{\psi_{\pm}}$ since $\theta^{2}=1$ for non-Kramers system. Hence, two eigenstates $\ket{\psi_{m}}$ and $\ket{\psi_{m'}}$ of $m^{\mathrm{th}}$ quasi-doublet can be chosen to have opposite parity with respect to the time-reversal symmetry, i.e.: 
\begin{gather}
\theta\ket{\psi_{m}}=\ket{\psi_{m}},\label{eq:phi_m}\\
\theta\ket{\psi_{m'}}=-\ket{\psi_{m'}}.\label{eq:phi_m'}
\end{gather}
 Here without loss of generality, we have assumed that $\ket{\psi_{m}}$, $\ket{\psi_{m'}}$ have parity 1 and $-1$ respectively (in case they have respectively parity $-1$ and 1, just simply switch the definition of $\ket{\psi_{m}}$ and $\ket{\psi_{m'}}$). Following the same way as in Ref. [\citenum{Abragam1970}], it is easy to prove that:
\begin{gather}
\braket{\psi_{m}|S_{\alpha}|\psi_{m}}=\braket{\psi_{m'}|S_{\alpha}|\psi_{m'}}=0,\\
\braket{\psi_{m}|S_{\alpha}|\psi_{m'}}=\braket{\psi_{m}|S_{\alpha}|\psi_{m'}}^{*}\equiv s_{m,\alpha}\in\Re,
\end{gather}
where $s_{m,\alpha}$ is real and $\alpha=x,y,z$. Remember that $s_{m,\alpha}$ is not necessarily an integer number for the molecular spin system.

We now find the localized basis states $\ket{m^{*}}$ which optimize the expectation value of $S_{z}$ by looking for $\alpha$ and $\varphi$ in the general form of the wave function $\ket{m^{*}}$: 
\begin{equation}
\ket{m^{*}}=\cos\alpha\ket{\psi_{m}}+e^{i\varphi}\sin\alpha\ket{\psi_{m'}}.
\end{equation}
Expectation value of $S_{z}$ then is: 
\begin{equation}
\braket{m^{*}|S_{z}|m^{*}}=\sin2\alpha\cos\varphi s_{m}.
\end{equation}
This quantity gets two extrema $\pm s_{m}$ at $\left(\alpha,\varphi\right)=\left(\pi/4,0\right)$ and $\left(\alpha,\varphi\right)=\left(\pi/4,\pi\right)$, which gives us two orthogonal localized basis states $\ket{\uparrow_{m}}$ and $\ket{\downarrow_{m}}$: 
\begin{gather}
\ket{\uparrow_{m}}=\frac{1}{\sqrt{2}}\left(\ket{\psi_{m}}+\ket{\psi_{m'}}\right),\\
\ket{\downarrow_{m}}=\frac{1}{\sqrt{2}}\left(\ket{\psi_{m}}-\ket{\psi_{m'}}\right).
\end{gather}
It should be emphasized that $\ket{\psi_{m}}$ and $\ket{\psi_{m'}}$ must be chosen with time-reversal symmetry following Eqs. (\ref{eq:phi_m}-\ref{eq:phi_m'}).  This is the key difference between our approach and previously mentioned approaches, and it also decides the uniqueness of the localized basis determination. 

It is easy to see that these two localized basis states are Kramers conjugates, i.e. $\ket{\downarrow_{m}}=\theta\ket{\uparrow_{m}}$ and $\ket{\uparrow_{m}}=\theta\ket{\downarrow_{m}}$. Consequently, we have, 
\begin{gather}
\braket{\uparrow_{m}|S_{\alpha}|\downarrow_{m}}=\braket{\downarrow_{m}|S_{\alpha}|\uparrow_{m}}=0,\,\alpha=x,y,z,\label{eq:Samm1}\\
\braket{\uparrow_{m}|S_{\alpha}|\uparrow_{m}}=-\braket{\downarrow_{m}|S_{\alpha}|\downarrow_{m}}=s_{m,\alpha},\label{eq:Samm2}\\
\braket{\uparrow_{m}|\hmt_{0}|\uparrow_{m}}=\braket{\downarrow_{m}|\hmt_{0}|\downarrow_{m}}=\varepsilon_{m},\\
\braket{\uparrow_{m}|\hmt_{0}|\downarrow_{m}}=\braket{\downarrow_{m}|\hmt_{0}|\uparrow_{m}}=\Delta_{m}/2.
\end{gather}

Interestingly, the above results show that two localized basis states corresponding to the doublet found above are also eigenstates of the operators $S_{\alpha}$, and $S_{z}$ in particular, in the subspace of the doublet. Hence, given the $z$-axis of the whole multiplet, another alternative procedure to generate the localized basis states is to diagonalize the submatrix $S_{z}$ in the doublet subspace, then make sure both eigenvectors are time-reversal symmetric.

It should be noted that the above determination of the localized basis where there exists a time-reversal symmetry between $\ket{\uparrow_{m}}$ and $\ket{\downarrow_{m}}$ of the $m^{\mathrm{th}}$ doublet allows us to write the spin Hamiltonian in the well-known form: 
\begin{equation}
\hmt_{0}=\sum_{m^{\mathrm{th}}}\varepsilon_{m}\left(\ket{\uparrow_{m}}\bra{\uparrow_{m}}+\ket{\downarrow_{m}}\bra{\downarrow_{m}}\right)+\sum_{m^{\mathrm{th}}}\frac{\Delta_{m}}{2}\left(\ket{\uparrow_{m}}\bra{\downarrow_{m}}+\ket{\downarrow_{m}}\bra{\uparrow_{m}}\right)+\sum_{n^{\mathrm{th}}}\varepsilon_{n}\ket{n}\bra{n}.
\end{equation}
In addition, Eq. \eqref{eq:Samm1} and \eqref{eq:Samm2} also implies that the Zeeman interaction cannot induce any additional off-diagonal term (tunnel splitting) in the doublet subspace.

\subsection{Kramers system}

For this kind of system, the doublet is fully degenerate. After diagonalizing the spin Hamiltonian $\hmt_{0}$, two orthogonal and time-reversal symmetric eigenstates, $\ket{\psi_{m}}$ and $\ket{\bar{\psi}_{m}}=\theta\ket{\psi_{m}}$, are obtained. Similar to the non-Kramers case, we suppose a general form of the localized basis states: 
\begin{equation}
\ket{m^{*}}=\cos\alpha\ket{\psi_{m}}+e^{i\phi}\sin\alpha\ket{\bar{\psi}_{m}},
\end{equation}
and then try to optimize the expectation value of $S_{z}$ in this subspace. Straightforward calculation gives 
\begin{equation}
\braket{m^{*}|S_{z}|m^{*}}=\cos2\alpha\braket{\psi_{m}|S_{z}|\psi_{m}}+\frac{1}{2}\sin2\alpha\left(e^{-i\varphi}\braket{\bar{\psi}_{m'}|S_{z}|\psi_{m}}+e^{i\varphi}\braket{\psi_{m}|S_{z}|\bar{\psi}_{m'}}\right).\label{eq:mSzm}
\end{equation}

The second term on the right-hand side indicates that transforming $\left\{ \ket{\psi_{m}},\ket{\bar{\psi}_{m}}\right\} $ to the eigenstates basis of $S_{z}$ is more convenient since this term will vanish. Continuing to diagonalize the $S_{z}$ matrix in the subspace $\left\{ \ket{\psi_{m}},\ket{\bar{\psi}_{m}}\right\} $, we obtain two new states, $\ket{\uparrow_{m}}$ and $\ket{\downarrow_{m}}=\theta\ket{\uparrow_{m}}$. These two are eigenstates of both $S_{z}$ in the doublet subspace and $\hmt_{0}$. That is to say: 
\begin{gather}
\hmt_{0}\ket{\uparrow_{m}}=\varepsilon_{m}\ket{\uparrow_{m}},\hmt_{0}\ket{\downarrow_{m}}=\varepsilon_{m}\ket{\downarrow_{m}},\\
\braket{\uparrow_{m}|S_{z}|\uparrow_{m}}=s_{m,},\braket{\downarrow_{m}|S_{z}|\downarrow_{m}}=-s_{m},\\
\braket{\uparrow_{m}|S_{z}|\downarrow_{m}}=\braket{\downarrow_{m}|S_{z}|\uparrow_{m}}=0.
\end{gather}
It should be reminded that $s_{m}$ is not necessarily a half-integer number in the case of the molecular spin system. After changing the basis from $\left\{ \ket{\psi_{m}},\ket{\psi_{m'}}\right\} $ to $\left\{ \ket{\uparrow_{m}},\ket{\downarrow_{m}}\right\} $, Eq. \eqref{eq:mSzm} becomes:
\begin{equation}
\braket{m^{*}|S_{z}|m^{*}}=\cos2\alpha\braket{\uparrow_{m}|S_{z}|\uparrow_{m}}=s_{m}\cos2\alpha.
\end{equation}
Obviously, $\braket{m^{*}|S_{z}|m^{*}}$ reaches two extrema $\pm s_{m}$ at respectively $\alpha=0$ and $\alpha=\pi/2$ regardless value of $\varphi$. In other words, two eigenstates of $S_{z}$, $\ket{\uparrow_{m}}$ and $\ket{\downarrow_{m}}$, optimize the expectation value of $S_{z}$ in the doublet subspace. They are thus localized basis states corresponding to the doublet we are searching.  

At first glance, the determination of the localized basis states for a Kramers doublet presented here is quite similar to the second approach mentioned previously \citep{Aquilante2016}. However, the \emph{key} difference in our process is that we only use one $z$-direction, which is along the main magnetic axis $z$ of the \emph{whole} spin multiplet for every doublet, instead of along magnetic axis $z$ of any specific doublet as in the second approach. Therefore, the optimization of the expectation value $\left\langle S_{z}\right\rangle $ for one doublet does not conflict with other doublets. 

\begin{figure}
\centering{}%
\begin{tabular}{cc}
\includegraphics[width=0.45\textwidth]{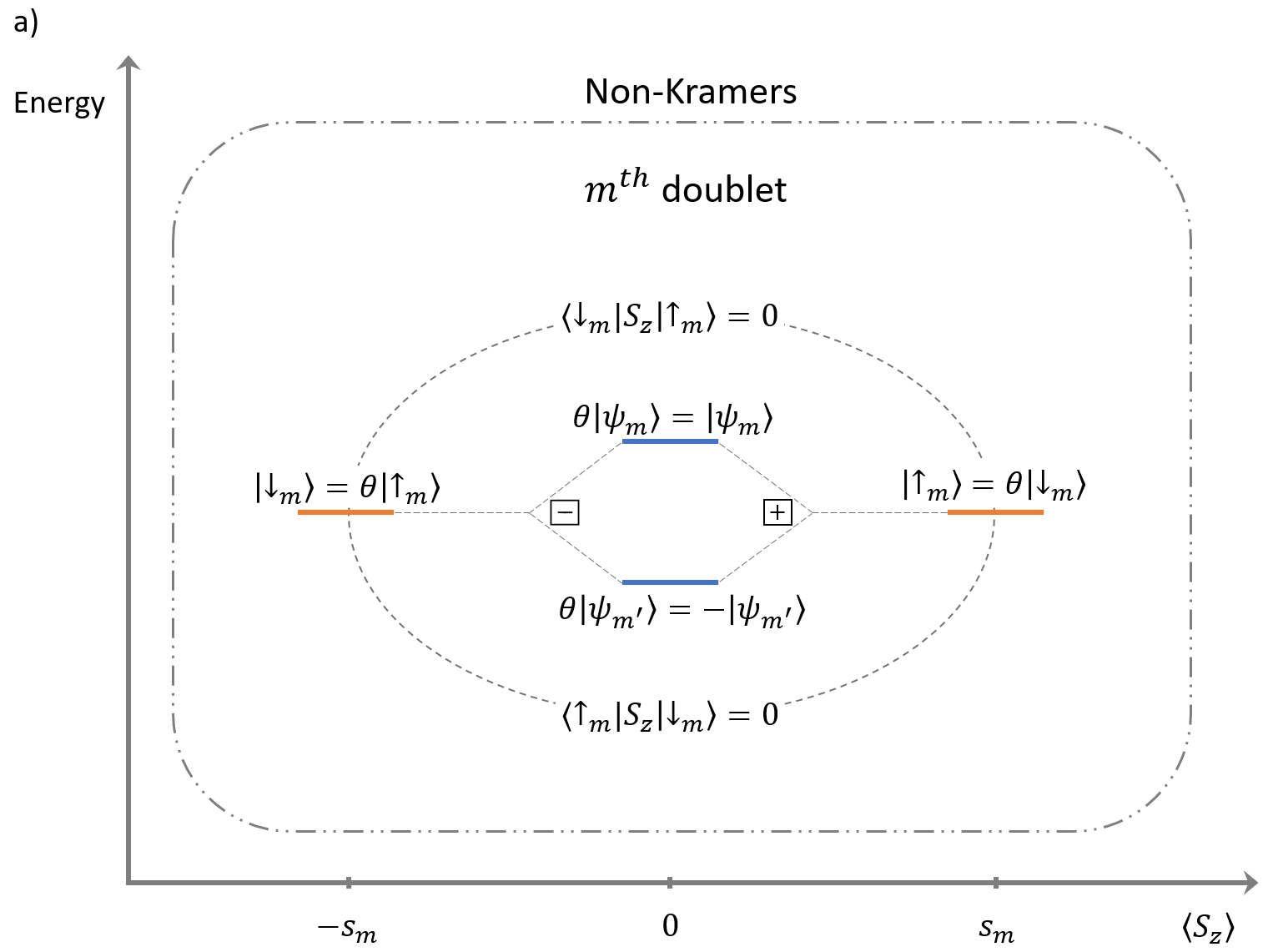} & \includegraphics[width=0.45\textwidth]{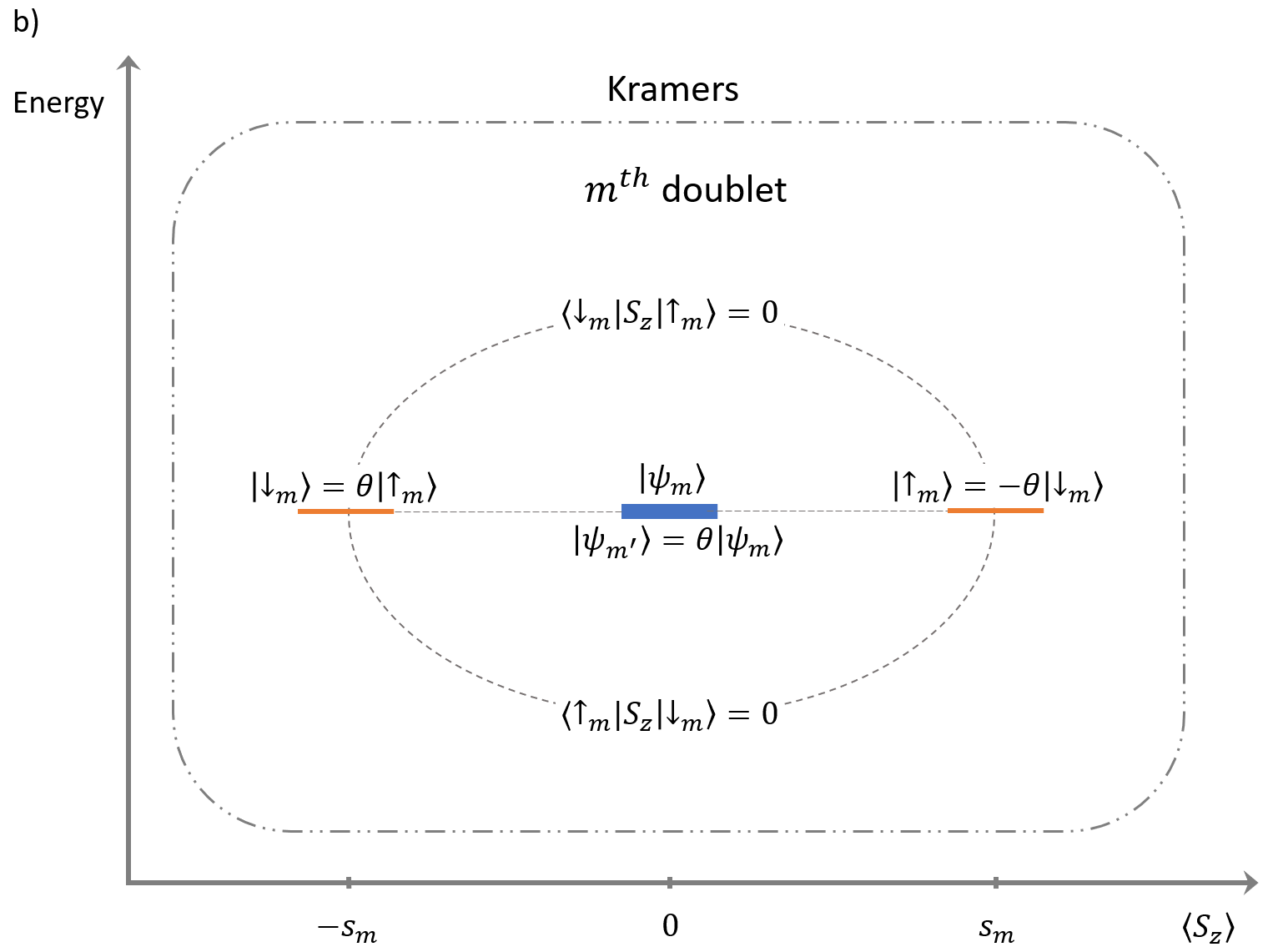}\tabularnewline
\end{tabular}\caption{Diagrams summarizing the procedures to obtain localized basis states $\ket{\downarrow_{m}}$ and $\ket{\uparrow_{m}}$ corresponding to the $m^{\mathrm{th}}$ doublet a) from the eigenstates $\ket{\psi_{m}}$ and $\ket{\psi_{m'}}$ which have opposite parity in non-Kramers system; b) from eigenstates $\ket{\psi_{m}}$ and its time-reversal symmetric $\ket{\psi_{m'}}=\theta\ket{\psi_{m}}$ in Kramers system. \label{fig:localized basis determination procedures}}
\end{figure}

We summarize the procedures to obtain localized basis states $\ket{\downarrow_{m}}$ and $\ket{\uparrow_{m}}$ corresponding to the $m^{\mathrm{th}}$ doublet for both non-Kramers and Kramers system in Fig. \ref{fig:localized basis determination procedures}.

\subsection{Example}

To exemplify the distinction between our approach and previous ones, a simple example featuring a spin $S=3/2$ is taken. In this example, all ESO coefficients $B_{pq}$ of the associated spin Hamiltonian are zero, except for $B_{2}^{0}=-B_{2}^{2}=-1$ (in arbitrary units). As usual, we align the $z$-axis with the primary magnetic axis of the whole spin. Diagonalizing this spin Hamiltonian yields the following eigenstates:
\begin{align}
\ket{1} & =-\alpha\ket{3/2}+\beta\ket{-1/2},\label{eq:ket1}\\
\ket{2} & =-\alpha\ket{-3/2}+\beta\ket{1/2}=\theta\ket{1},\\
\ket{3} & =\beta\ket{-3/2}+\alpha\ket{1/2},\\
\ket{4} & =-\beta\ket{3/2}-\alpha\ket{-1/2}=\theta\ket{3},\label{eq:ket4}
\end{align}
where $\alpha=\sqrt{2+\sqrt{3}}/2$, $\beta=\sqrt{2-\sqrt{3}}/2$, and the corresponding eigenenergies $\varepsilon_{1}=\varepsilon_{2}=-\varepsilon_{3}=-\varepsilon_{4}=-2\sqrt{3}$.  

Within our approach, it is straightforward to confirm that the above eigenstates $\left\{ \ket{1},\ket{2}\right\} $ and $\left\{ \ket{3},\ket{4}\right\} $ can be taken as localized basis corresponding to the ground and excited doublet, respectively. That is to say, they optimize the expectation value of $S_{z}$ in each doublet subspace. Indeed, a calculation of the $S_{z}$ matrix in this basis results in: 
\begin{equation}
S_{z}=\frac{1}{2}\left(\begin{array}{cccc}
\sqrt{3}+1 & 0 & 0 & 1\\
0 & -\sqrt{3}-1 & 1 & 0\\
0 & 1 & \sqrt{3}-1 & 0\\
1 & 0 & 0 & 1-\sqrt{3}
\end{array}\right).
\end{equation}
As can be seen, the sub-matrices of $S_{z}$ corresponding to each doublet in this basis are diagonal. Due to the freedom arising from the Kramers degeneracy, others may choose a different set of initial eigenstates which may not optimize the expectation value of $S_{z}$. However, following our procedure given in previous section, the same results can be easily obtained after diagonalizing sub-matrices of $S_{z}$ corresponding to each doublet.

Let's attempt to follow the guidelines to obtain the localized basis using the previous first approach \citep{Gatteschi2007}. From the expressions Eqs. (\ref{eq:ket1}-\ref{eq:ket4}), it is obvious that if we sum or subtract any two eigenstates belonging to any doublet without considering their phase factor and their time-reversal symmetry, or adding a phase factor to any of them, as in the procedure given by the previous first approach \citep{Gatteschi2007}, this will certainly result in somewhat \emph{delocalized }or not localization-optimized states. For example, if we take $\ket{\psi_{\pm}}=\left(\ket{1}\pm\ket{2}\right)/\sqrt{2}$, expectation value of $S_{z}$ would then be $\braket{\psi_{\pm}|S_{z}|\psi_{\pm}}=0$. The resulting states $\ket{\psi_{\pm}}$ are thus completely delocalized and do not satisfy the objective of constructing a localized basis from eigenstates of the doublets.

Following the procedure given in the previous second approach \citep{Chibotaru2012b}, a calculation of the main magnetic axis $z$ of the ground and excited doublet when considered as pseudospins $\tilde{s}=1/2$  reveals that while the main magnetic axis $z$ of the ground doublet (pseudospin $\tilde{s}=1/2$) coincides with the chosen $z$-axis of the whole multiplet $S=3/2$, the main magnetic axis $z$ of the excited doublet (pseudospin $\tilde{s}=1/2$) is oriented along the $x$-axis, i.e. perpendicular to the main magnetic axis $z$ of the ground doublet (pseudospin $\tilde{s}=1/2$). Applying a magnetic field along each doublet's main magnetic axis $z$ \citep{Chibotaru2012b,Ungur2011a,Aquilante2016} then results in the following basis states with corresponding $S_{z}$ expectation value: 
\begin{align}
\ket{1'} & =\ket{1},\left\langle S_{z}\right\rangle =\left(\sqrt{3}+1\right)/2,\label{eq:ket1'}\\
\ket{2'} & =\ket{2},\left\langle S_{z}\right\rangle =-\left(\sqrt{3}+1\right)/2,\\
\ket{3'} & =\left(\ket{4}+\ket{3}\right)/\sqrt{2},\left\langle S_{z}\right\rangle =0,\\
\ket{4'} & =\left(\ket{4}-\ket{3}\right)/\sqrt{2},\left\langle S_{z}\right\rangle =0.\label{eq:ket4'}
\end{align}
As can be seen, if we take these four states as the localized basis, the states $\ket{3'}$ and $\ket{4'}$ will have a delocalized magnetic moment. 

To highlight the significance of choosing an appropriate localized basis in molecular magnetism, let's consider the bases presented above, Eqs. (\ref{eq:ket1}-\ref{eq:ket4}) and Eqs. (\ref{eq:ket1'}-\ref{eq:ket4'}). We will employ these bases to interpret magnetization relaxation and show that when utilizing the second basis Eqs. (\ref{eq:ket1'}-\ref{eq:ket4'}), if the interpretation of the relaxation process is not approached carefully, it may result in an incorrect explanation of the magnetization relaxation mechanism and the dominant relaxation pathway in the molecular spin. 

\begin{figure}
\centering{}%
\begin{tabular}{cc}
\includegraphics[width=0.45\textwidth]{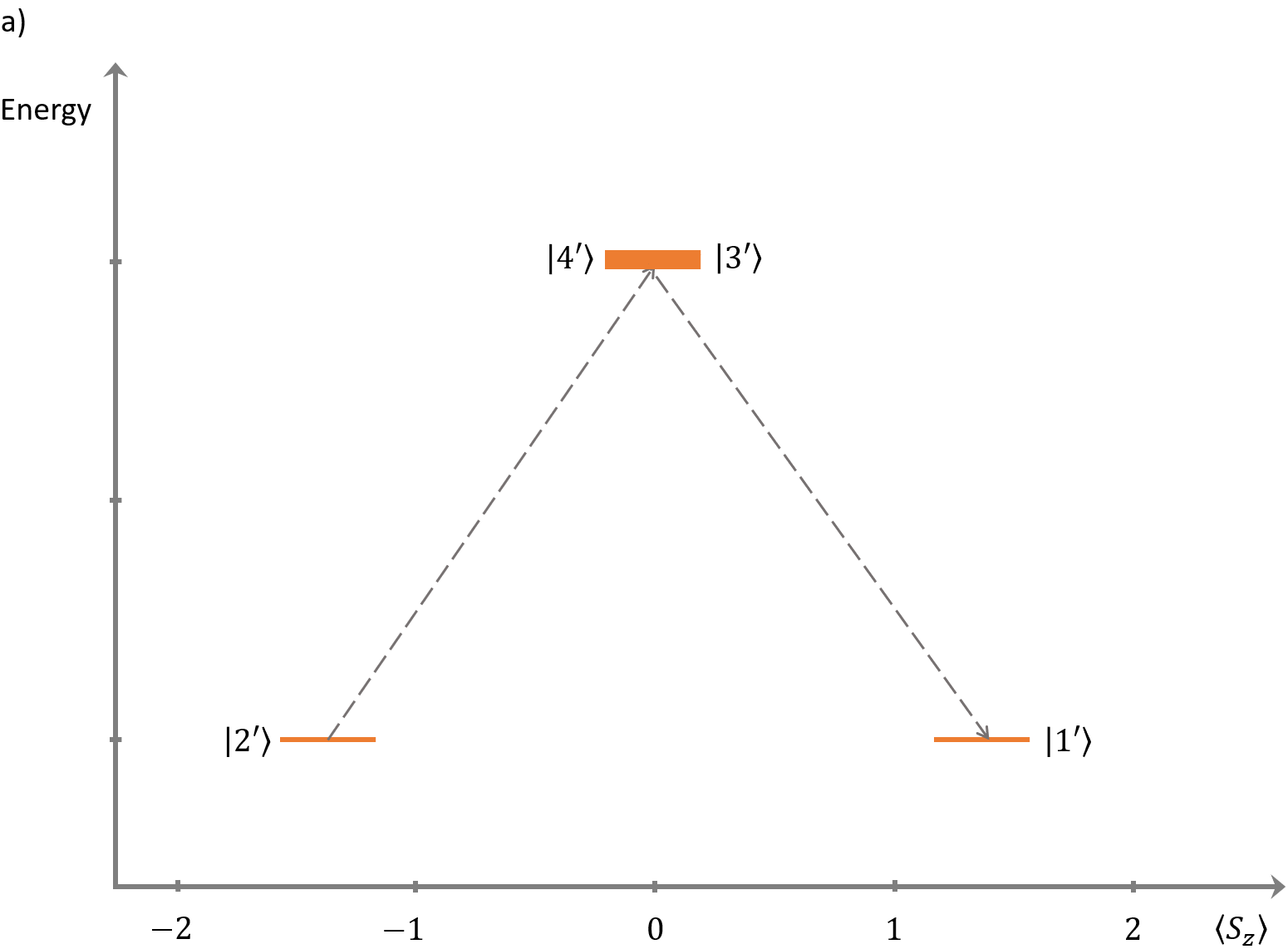} & \includegraphics[width=0.45\textwidth]{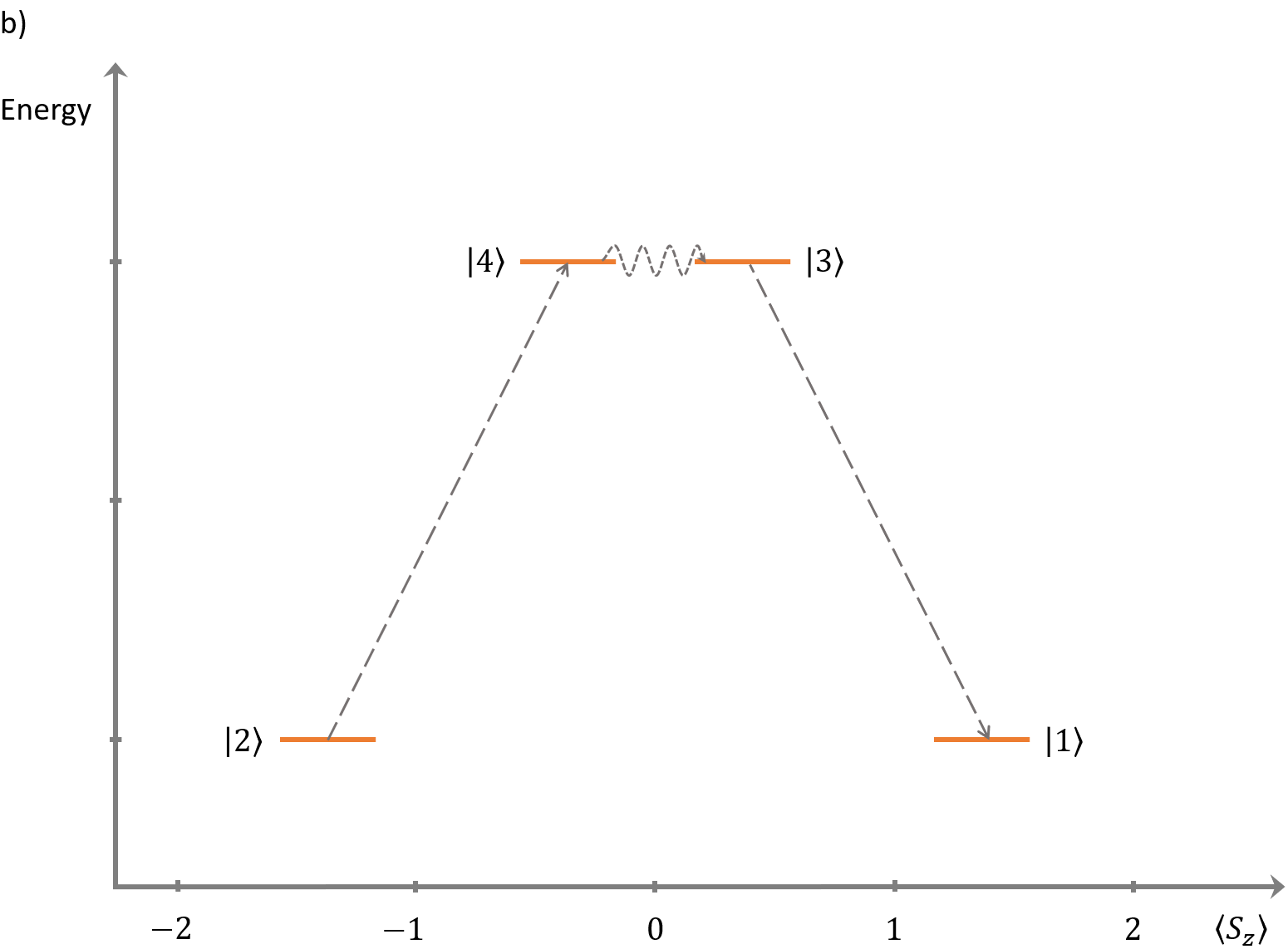}\tabularnewline
\end{tabular}\caption{Dominant relaxation path diagram with localized basis states extracted a) by using the determination procedures from the previous second approach \citep{Chibotaru2012b,Ungur2011a,Aquilante2016} and b) by using the determination procedures proposed in this work. \label{fig:example}}
\end{figure}

In Fig. \ref{fig:example}, we provide diagrams depicting the dominant magnetization relaxation pathways obtained through the determination of the localized basis using the previously mentioned second approach \citep{Chibotaru2012b,Ungur2011a,Aquilante2016} and our approach. Examining Fig. \ref{fig:example}a, it becomes evident that using the determination procedures from the former approach, one might erroneously attribute an Orbach process from $\ket{2'}$ via $\ket{4'}$ and $\ket{3'}$ toward $\ket{1'}$ as the primary magnetization relaxation mechanism. However, this does not align with the actual mechanism, which involves the crucial role of quantum tunneling of magnetization.

On the other hand, using the localized basis determined with our new approach, as shown in Fig. \ref{fig:example}b, allows for a straightforward interpretation of magnetization relaxation, where the transition proceeds from state $\ket{2}$ via a direct pathway to $\ket{4}$, followed by quantum tunneling to $\ket{3}$, and ultimately emitting to $\ket{1}$. Clearly, the revised localized basis determination process simplifies the interpretation of the primary relaxation mechanism and provides a comprehensible physics-based perspective.

It should be noticed that the delocalized issue in the determination of the localized basis given by the previous second approach is magnified when there is a significant difference in orientation between the main magnetic axis $z$ of the whole spin and each individual doublet, as well as between different doublets. This can be observed in the aforementioned example, where the ground and first excited doublets have perpendicular main magnetic axes $z$. 

\section{Discussions}

In both theoretical models and \emph{ab initio} calculations, it is not uncommon to encounter triplets and quadruplets alongside doublets. The procedures for determining the localized basis in these situations can be readily generalized from the doublet case, applicable to both non-Kramers and Kramers molecular spin systems. This generalization is rooted in the earlier findings, which establish that the localized basis states associated with a doublet are eigenstates of the $S_{z}$ operator with opposite parity and reside in the subspace spanned by the doublet.

In the case of a triplet, diagonalizing the $S_{z}$ operator (or equivalently, the longitudinal magnetic moment $\mu_{z}$ along the main magnetic axis $z$ of the whole spin system) in the triplet's subspace will result in three eigenvectors and three eigenvalues. Among them, two eigenvalues will have opposite values, and their corresponding eigenvectors are time-reversal symmetric. The doublet subspace can thus be constructed from these two eigenvectors. The remaining distinct eigenvalue/eigenvector must belong to the singlet state. Consequently, the triplet subspace is fully decomposed, and we obtain the corresponding localized basis states.

In the case of a quadruplet, which essentially comprise two degenerate doublets, the same process can be employed to separate the two doublet subspaces and determine their corresponding localized basis states/subspaces.

In summary, this work has clearly demonstrated that a unique determination of localized basis states in molecular spins can be achieved by optimizing the longitudinal spin component along the main magnetic axis $z$ of the whole multiplet of the molecular spin. This optimization is grounded in time-reversal symmetry and $S_{z}$ operator along the main magnetic axis $z$ of the molecular spin. It allows us to associate a wave function (localized basis state) with either an ``up'' or ``down'' magnetic moment orientation within a molecular spin doublet, irrespective of the initial form of its energy eigenstates. Our finding thus facilitates a comprehensible interpretations of magnetic dynamics, such as the relaxation or decoherence processes, in molecular spins. It also holds great potential for aiding in the development of simplified theoretical models involving these materials. The proposed determination method can be readily employed in conjunction with existing quantum chemistry packages \citep{Aquilante2016a,Chibotaru2012b,Ungur2011a,Neese2012,Neese2018a,neese2022} to qualitatively estimate the dominant relaxation path in single-molecule magnets or any future quantitative methods with a similar objective. 
\begin{acknowledgments}
L. T. A. H. and L. U. acknowledge the financial support of the research projects R-143-000-A65-133, A-8000709-00-00, and A-8000017-00-00 of the National University of Singapore. Calculations were done on the ASPIRE-1 cluster (www.nscc.sg) under the projects 11001278 and 51000267. Computational resources of the HPC-NUS are gratefully acknowledged.
\end{acknowledgments}

\bibliographystyle{apsrev4-2}
\bibliography{library}

\end{document}